\documentclass[%
 aip,
 amsmath,amssymb,
 reprint,%
]{revtex4-2}

\usepackage{graphicx}% Include figure files
\usepackage{dcolumn}% Align table columns on decimal point
\usepackage{bm}% bold math
\usepackage[utf8]{inputenc}
\usepackage[T1]{fontenc}
\usepackage{mathptmx}
\usepackage{etoolbox}

\usepackage{twoopt}
\usepackage{mathtools}
\newcommand{\rem}[1]{ } % Comment text out

\newcommand{\f}{{\cal F}}
\newcommand{\g}{{\cal G}}
\newcommand{\del}{\nabla}
\newcommand{\ee}{{\bf E}}
\newcommand{\bb}{{\bf B}}
\newcommand{\F}{\left(\frac{\bb^{2}-\ee^{2}}{2}\right)}
\newcommand{\G}{\left(-\bb\cdot\ee\right)}
\newcommand{\cdt}{\frac{1}{c}\frac{\partial}{\partial t}}

\newcommand{\araa}{{Ann. Rev. Astron. Astroph.}}
\newcommand{\mnras}{{Mon. Not. R. Astron. Soc.}}
\makeatletter
\def\@email#1#2{%
 \endgroup
 \patchcmd{\titleblock@produce}
  {\frontmatter@RRAPformat}
  {\frontmatter@RRAPformat{\produce@RRAP{*#1\href{mailto:#2}{#2}}}\frontmatter@RRAPformat}
  {}{}
}%
\makeatother

\begin{document}

%\preprint{AIP/123-QED}

\title[QED plasma modes]{Plasma modes in QED super-strong magnetic fields of magnetars and laser plasmas}
% Force line breaks with \\
\author{Mikhail V. Medvedev}\thanks{Corresponding author.}
\email{medvedev@ku.edu}
% \altaffiliation[Also at ]{Laboratory for Nuclear Science, Massachusetts Institute of Technology, Cambridge, MA 02139}%Lines break automatically or can be forced with \\
%\author{B. Author}%
% \email{Second.Author@institution.edu.}
\affiliation{Institute for Advanced Study, School for Natural Sciences, Princeton, NJ 08540}%
\affiliation{Department of Astrophysical Sciences, Princeton University, Princeton, NJ 08544}%
\affiliation{Department of Physics and Astronomy, University of Kansas, Lawrence, KS 66045
%\\This line break forced with \textbackslash\textbackslash
}%
\affiliation{Laboratory for Nuclear Science, Massachusetts Institute of Technology, Cambridge, MA 02139
%\\This line break forced with \textbackslash\textbackslash
}%

% \homepage{http://www.Second.institution.edu/~Charlie.Author.}
%\affiliation{%
%Second institution and/or address%\\This line break forced% with \\
%
%}%


\begin{abstract}
Ultra-magnetized plasmas, where the magnetic field strength exceeds the Schwinger field of about $B_{Q}\approx4\times10^{13}$~gauss, become of great scientific interest, thanks to the current advances in laser-plasma experiments and astrophysical observations of magnetar emission. These advances demand better understanding of how quantum electrodynamics (QED) effects influence collective plasma phenomena. In particular, Maxwell's equations become nonlinear in the strong-QED regime. Here we present the `QED plasma framework' which will allow one to {\em systematically} explore collective phenomena in a QED-plasma with arbitrarily strong magnetic field. Further, we illustrate the framework by exploring low-frequency modes in the ultra-magnetized, cold, electron-positron plasmas. We demonstrate that the classical picture of five branches holds in the QED regime; no new eigenmodes appear. The dispersion curves of all the modes are modified. The  QED effects include the overall modification to the plasma frequency, which becomes field-dependent. They also modify resonances and cutoffs of the modes, which become both field- and angle-dependent. The strongest effects are (i) the {\em field-induced transparency of plasma} for the O-mode via the dramatic reduction of the low-frequency cutoff well below the plasma frequency, (ii) the {\em Alfven mode suppression} in the large-$k$ regime via the reduction of the Alfven mode resonance, and (iii) the {\em O-mode slowdown} via strong angle-dependent increase of the index of refraction. These results should be important for understanding of a magnetospheric pair plasma of a magnetar and for laboratory laser-plasma experiments in the QED regime.
\end{abstract}

\maketitle

%%%%%%%%%%%%%%%%%%%%%%%%%%%%%%%%%%%%%%%%%%%%

\section{Introduction}
\label{s:intro}

Systems in which electromagnetic fields approach or even exceed the Quantum Electrodynamic (QED) strength, $B_{Q}=m_{e}^{2}c^{3}/\hbar e\simeq 4\times 10^{13}$~gauss, exist in nature. Magnetars --- the strongly magnetized neutron stars --- have surface magnetic fields of about $10^{14}-10^{15}$~gauss \cite{magnetars-rev}. The `central engines' of gamma-ray bursts are expected to possess even stronger magnetic fields, up to $10^{15}-10^{17}$~gauss \cite{grb-magnetar-fields}. The laboratory multi-petawatt lasers are rapidly approaching the quantum limit too \cite{qed-lasers-rev}. Propagation of electromagnetic and MHD waves in an extreme QED regime has been studied earlier \cite{DGbook,nl-qed-heyl,mhd-qed-extreme}. It was shown, for example, that  vacuum becomes bi-refringent, so that the index of refraction is polarization-dependent. The effects of three-wave photon splitting, the scattering of light off light and off the ambient $B$-field in vacuum have also been shown \cite{photon-split-adler,scatt-light-light}. Recently, the QED effects were included in computational codes and are used, for instance, in simulations of magnetic reconnection in plasmas with QED fields and the generation of QED cascades in laser plasmas \cite{qed-solver,qed-cascades-laser-17,qed-reconn-19,qed-reconn-23}.

It is important to note, however, that the electromagnetic fields in magnetars and laser experiments are not vacuum fields. In contrast, the systems have a substantial plasma component. Thus, the electromagnetic fields are produced by plasma collective phenomena, such as waves and instabilities.
In magnetars, the magnetosphere can twisted by surface shear motions, so it is threaded by electric currents ${\bf j}=(c/4\pi)\nabla\times{\bf B}$. Thus, their magnetospheres carry electron-positron plasma needed to maintain the current, $n=j/c e\simeq 10^{17} B_{15}r_{6}^{-1}{\rm cm}^{-3}$, where $B_{15}=B/(10^{15}\textrm{ gauss})$ and $r_{6}=r/(10^{6}\textrm{ cm})$ is the radial distance from the center of the magnetar. In laser experiments, plasma is either created by the interaction with a target, or the $e^{\pm}$ pair plasma can be created from in the laser beam via the Breit-Wheeler process \cite{positron-from-laser-exp,breight-wheeler-plasma-from-light}. The number density of the laser plasma can be high, up to a few~$\times10^{19}{\rm cm}^{-3}$. Thus, these systems naturally explore the very interesting regime where the plasma collective phenomena and strong-field quantum effects intertwine. It is thus imperative to develop the `QED plasma framework', which couples Nonlinear Maxwell's equations with plasma dynamics. It will allow one to explore this QED-plasma  domain via theoretical, computational, and experimental methods. 

In general, the studies of the vacuum polarization by strong fields have a long history of many decades. Some formalism described in the beginning of this paper appears in earlier works. In particular, references \cite{HL06,P+04} nicely summarize earlier findings and contain various analytical expressions and fitting formulae. Here we briefly present the relevant results for completeness and extent them to the case of supercritical fields. Further, we use them to explore plasma modes in such a field.

The rest of the paper is organized as follows. Section \ref{s:basic} outlines the theory of QED vacuum corrections to Maxwell's equations. Section \ref{s:framework} presents the QED-plasma framework, Section \ref{s:analysis} presents analytical results on QED-plasma eigenmodes in a cold electron-position plasma and Section \ref{s:results} illustrates them via the full numerical solution of the dispersion equation. Section \ref{s:concl} summarizes the main outcomes of the presented study.

\section{Basic theory}
\label{s:basic}

\subsection{Introduction}

In classical electrodynamics, vacuum is defined as an empty space. Hence it is a passive background where particles and fields propagate and evolve. Maxwell's equations are linear in their fields and are sourced by charges and currents. When the electromagnetic fields become strong enough, QED effects become important and make Maxwell's equations nonlinear. The nonlinearity arises from the polarization of quantum vacuum. 

Let's consider the effect of the electric field first. From the Heisenberg uncertainty principle $\Delta p \Delta x \sim \hbar$, one can estimate the scale on which a virtual electron-positron ($e^{\pm}$) pair exists to be $\lambda_{e} \sim \hbar/m_{e}c\sim10^{-11}{\rm cm}$. If the electric field of strength $E$ is present, it does work on the particles. Each virtual particle gains energy ${\cal E}\sim e E \lambda_{e}=e E \hbar/m_{e}c^{2}$ and if ${\cal E}\gtrsim m_{e}c^{2}$ the virtual pair becomes real. This `vacuum breakdown' occurs when the electric filed strength exceeds the critical (or `quantum', or Schwinger) field $E_{Q}=m_{e}^{2}c^{3}/\hbar e$. 

Now, let's consider magnetic field. A nonrelativistic charged particle executes Larmor gyration in magnetic field (in the perpendicular plane) with the frequency $\Omega=e B/m_{e}c$. From the uncertainly principle $\Delta {\cal E} \Delta t\sim \hbar$, we can write $\Delta {\cal E}\sim \hbar /\Delta t \sim \hbar \Omega$. The field in which $\Delta {\cal E}\sim m_{e}c^{2}$ corresponds to the quantum field $B_{Q}=m_{e}^{2}c^{3}/\hbar e\simeq 4\times 10^{13}$~gauss. Thus, we see that when the electromagnetic field strength approached the quantum field strength, QED effects cannot be neglected.

\subsection{Nonlinear Maxwell's equations}

The set of electromagnetic field equations is obtained from the variational principle for action $\delta S=\delta \int {\cal L}\ dVdt=0$. The classical Lagrange density ${\cal L}={\cal L}_{f}+{\cal L}_{fm}$ for the fields and field-matter interaction yields the classical Maxwell's equations, where
\begin{eqnarray}
{\cal L}_{f} & = & -\frac{1}{16\pi}F_{\mu\nu}F^{\mu\nu}=\frac{1}{8\pi}\left({\bf E}^2-{\bf B}^2\right),\\
{\cal L}_{fm} & = & -\frac{1}{c}A_{\mu}j^{\mu},
\end{eqnarray}
$F_{\mu\nu}=\partial_{\mu}A_{\nu}-\partial_{\mu}A_{\nu}$ is the Faraday tensor, $A_{\mu}$ is the electromagnetic 4-potential and $j^{\mu}$ is the 4-current. Hereafter, summation is assumed over repeated (`dummy') indices. Greek indices run from 0 throught 3 and roman indices run from 1 through 3.

The QED correction in the one-loop approximation to the classical Lagrangian is given by \citep{EK35, HE36, W36}
\begin{align}
{\cal L}'_{f} &=\frac{m_{e}c^2}{8\pi^2}
\left(\frac{m_{e}c}{\hbar}\right)^3 
\int_0^\infty\frac{e^{-\eta}}{\eta^3} 
\nonumber\\
&\quad\times
\left[-\left(\eta a\cot\eta a\right)\left(\eta b\coth\eta b\right)
+1-\frac{\eta^2}{3}\left(a^2-b^2\right)\right]d\eta,
\end{align}
that is the full Lagrangian density is ${\cal L}={\cal L}_{f}+{\cal L}'_{f}+{\cal L}_{fm}$. 
This approximation includes all one-loop diagrams with various numbers of vertexes, so it represents a first-order contribution in $\alpha$, the fine structure constant. Higher-order multi-loop diagrams are omitted. Here the parameters
\begin{align}
a&=\frac{\hbar e E}{m_{e}^2 c^3}\equiv\frac{E}{E_{Q}}, \\
b&=\frac{\hbar e B}{m_{e}^2 c^3} \equiv\frac{B}{B_{Q}}
\end{align}
can be written in a covariant form, so they are applicable in any reference frame,
\begin{align}
a&=-\frac{i\hbar e}{\sqrt{2}m_{e}^2 c^3}
\left[\left({\cal F}+i{\cal G}\right)^{1/2}-\left({\cal F}-i{\cal G}\right)^{1/2}\right]\\
b&=\frac{\hbar e}{\sqrt{2}m_{e}^2 c^3}
\left[\left({\cal F}+i{\cal G}\right)^{1/2}+\left({\cal F}-i{\cal G}\right)^{1/2}\right].
\end{align}
The invariants are
\begin{eqnarray}
{\cal F} &=& \frac{1}{4} F_{\mu\nu}F^{\mu\nu}=\frac{1}{2}\left({\bf B}^2-{\bf E}^2\right),\\
{\cal G} &=& \frac{1}{4} \hat F_{\mu\nu} \hat F^{\mu\nu}=-{\bf B}\cdot{\bf E},\\
{\cal F}\pm i{\cal G} &=& \frac{1}{2}\left({\bf B}^2\pm i{\bf E}^2\right),
\end{eqnarray}
where $\hat F^{\mu\nu}=\frac{1}{2}\varepsilon^{\mu\nu\gamma\delta}F_{\gamma\delta}$ is the Hodge dual tensor and $\varepsilon^{\mu\nu\gamma\delta}$ is the Levi-Civita symbol. 

The first pair of Maxwell's equations, $\partial_{\nu}\hat F^{\mu\nu}=0$, is unaffected by the QED corrections:
\begin{eqnarray}
\nabla\cdot {\bf B} &=&0,
\label{m1}\\
\frac{1}{c}\frac{\partial {\bf B}}{\partial t}+\nabla\times {\bf E} &=& 0.
\label{m2}
\end{eqnarray}

The second pair is obtained from the Euler-Lagrange equation
\begin{equation}
\frac{\partial {\cal L}}{\partial A_{\nu}}-\frac{\partial}{\partial x^{\mu}}\frac{\partial {\cal L}}{\partial A_{\nu,\mu}}=0.
\end{equation}
These equations can be written as follows \citep{landavshitz}
\begin{eqnarray}
\nabla\cdot {\bf D} &=& 4\pi \rho,
\label{gauss}\\
 \frac{1}{c}\frac{\partial {\bf D}}{\partial t} -\nabla\times {\bf H} &=& -\frac{4\pi}{c} {\bf j},
\label{ampere}
\end{eqnarray}
where 
\begin{eqnarray}
{\bf D} &=& {\bf E} + 4\pi {\bf P(E,B)},
\label{d}\\
{\bf H} &=& {\bf B} - 4\pi {\bf M(E,B)}. 
\label{h}
\end{eqnarray}
The QED-induced polarization and magnetization vectors are
\begin{align}
{\bf P(E,B)} &= \frac{\partial {\cal L}'_{f}}{\partial {\bf E}},\\
{\bf M(E,B)} &= \frac{\partial {\cal L}'_{f}}{\partial {\bf B}}.
\end{align}
Since ${\bf P}$ and ${\bf M}$ are highly nonlinear functions of the fields, we see that QED effects make Maxwell's equations nonlinear in vacuum. 

The plasma contribution enters, as usual, via the current term in Ampere's law, Eq. (\ref{ampere}). The plasma response can be calculated as usual via a (frequency-dependent) conductivity, for example, relating ${\bf j}$ to ${\bf E}$. The detailed form of the plasma response depends on the problem at hand. QED effects also modify electrostatic effects in plasma (e.g., Debye shielding) via Gauss' law, Eq. (\ref{gauss}).
 
The general analytical equation for ${\bf P}$ and ${\bf M}$ is not known. Below we consider the most interesting case of an arbitrarily strong magnetic field with $E=0$.

\subsection{Explicit form of nonlinear Maxwell's equations}

To express Maxwell's equations in the arbitrarily strong magnetic fields it is more convenient to express them somewhat differently. Indeed, the equations of motion are obtained from the Euler-Lagrange equation with the Lagrangian density, ${\cal L}={\cal L}_{f}+{\cal L}'_{f}+{\cal L}_{fm}$, given above
\begin{equation}
\partial_{\mu}\frac{\partial{\cal L}}{\partial F_{\mu\nu}}=\frac{1}{2c}j^{\nu}.
\end{equation}
Using $\partial{\cal F}/\partial F_{\mu\nu}=F^{\mu\nu}/2$, $\partial{\cal G}/\partial F_{\mu\nu}={\hat F}^{\mu\nu}/2$ and Bianchi identity ${\hat F^{\mu\nu}}_{,\mu}=0$, the nonlinear QED-vacuum Maxwell's equations become \cite{LundinPHD,DGbook}
\begin{align}
\gamma_{\cal F}\partial_{\mu}F^{\mu\nu}
&+\frac{1}{2}\Bigl[\gamma_{\cal F F}F^{\mu\nu}F_{\alpha\beta}
+\gamma_{\cal G G}{\hat F}^{\mu\nu}{\hat F}_{\alpha\beta}
\Bigr.\nonumber\\
&\Bigl.
+\gamma_{\cal F G}\left(F^{\mu\nu}{\hat F}_{\alpha\beta}+{\hat F}^{\mu\nu}F_{\alpha\beta}\right)\Bigr]\partial_{\mu}F^{\alpha\beta}=\frac{1}{c}j^{\nu},
\end{align}
where $\gamma_{\f}=\partial{\cal L}/\partial{\cal F}$, $\gamma_{\f\f}=\partial^{2}{\cal L}/\partial{\f}^{2}$,
 $\gamma_{\g\g}=\partial^{2}{\cal L}/\partial{\g}^{2}$, $\gamma_{\f\g}=\partial^{2}{\cal L}/\partial{\f}\partial{\g}$. These equations can be expressed explicitly in term of the $\ee$ and $\bb$ fields, as follows.

Modified Gauss' law:
\begin{align}
&\gamma_{\f}\del\cdot{\ee}+\gamma_{\f\f}\ee\cdot\del\F+\gamma_{\g\g}\bb\cdot\del\G
\nonumber\\
&
+\gamma_{\f\g}\left[\ee\cdot\del\G+\bb\cdot\del\F\right]=-\rho,
\label{gauss-my}
\end{align}

Modified Ampere's law:
\begin{align}
&\gamma_{\f}\left[\cdt\ee-\del\times\bb\right]
\nonumber\\
&
+\gamma_{\f\f}\left[\ee\cdt\F+\bb\times\del\F\right]
\nonumber\\
&
+\gamma_{\g\g}\left[\bb\cdt\G-\ee\times\del\G\right]
\nonumber\\
&
+\gamma_{\f\g}\left[\ee\cdt\G+\bb\times\del\G\right.
\nonumber\\
&
\left.+\bb\cdt\F-\ee\times\del\F\right]=\frac{1}{c}{\bf j}.
\label{ampere-my}
\end{align}

In the last two equations, the first term on the left-hand-side with $\gamma_{\f}=-1/(4\pi)$ and the term on the right-hand-side represent the standard (linear) Maxwell's equations.

In the case of a non-zero $\bb$ and the vanishing $\ee$ fields, the $\gamma$ coupling scalars are evaluated to be
\begin{align}
\gamma_{\f} & = -(1- C_{\delta})/(4\pi),\\
\gamma_{\f\f} & = C_{\mu}/(4\pi B^{2}),\\
\gamma_{\g\g} & = C_{\epsilon}/(4\pi B^{2}),\\
\gamma_{\f\g} & = 0.
\end{align}

where \cite{LundinPHD,DGbook}
\begin{align}
C_{\delta}=&\alpha\left[4 \zeta^{(1,0)}\left(-1,\frac{1}{2 b}\right)-\frac{1}{4 b^2}+\frac{1}{3}\log (2 b)\right.
\nonumber\\
&\qquad \left.+\frac{1}{2b}\log \left(\frac{\pi }{b}\right)-\frac{1}{b}\log \left(\Gamma\left(\frac{1}{2 b}\right)\right)-\frac{1}{6}\right],
\label{cd}\\
C_{\epsilon}=&\alpha\left[4 \zeta^{(1,0)}\left(-1,\frac{1}{2 b}\right)-\frac{1}{3} \psi\left(1+\frac{1}{2 b}\right)-\frac{1}{4b^2}+\frac{b}{3}\right.
\nonumber\\
&\qquad \left.+\frac{1}{2 b}\log \left(\frac{\pi }{b}\right)-\frac{1}{b}\log \left(\Gamma \left(\frac{1}{2b}\right)\right)-\frac{1}{6}\right],
\label{ce}\\
C_{\mu}=&\alpha\left[\frac{1}{2 b^2}\psi\left(1+\frac{1}{2 b}\right)-\frac{1}{2 b^2}-\frac{1}{2b}\right.
\nonumber\\
&\qquad \left.+\frac{1}{2 b}\log (4 \pi  b)-\frac{1}{b}\log \left(\Gamma \left(\frac{1}{2b}\right)\right)+\frac{1}{3}\right].
\label{cm}
\end{align}
We remind that $b=B/B_{Q}$ and $\alpha\equiv e^2/\hbar c\approx1/137$ is the fine structure constant. Here $\zeta^{(1,0)}(s,x)\equiv\partial_{s}\zeta(s,x)$ is the first derivative of the Hurwitz zeta-function with respect to its first argument, $\Gamma(x)$ is the Gamma-function, $\psi(x)\equiv\psi^{(0)}(x)=\Gamma'(x)/\Gamma(x)=\partial_{x}\log\Gamma(x)$ is the polygamma function of zeroth order being the logarithmic derivative of the Gamma-function. 

\begin{figure}
\centering
\includegraphics[scale = 0.9]{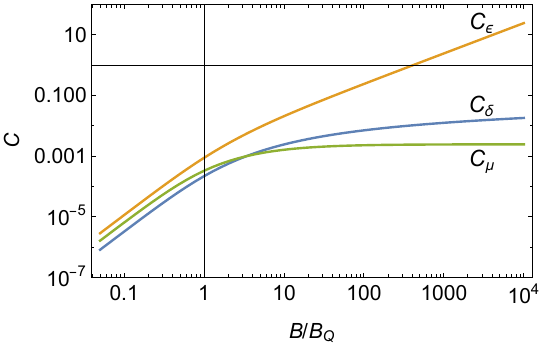}  %% file name without extension
\caption[]{Quantities $C_{\delta}, C_{\epsilon}, C_{\mu}$ as functions of $B/B_{Q}$.}
\label{f:C}
\end{figure}

Note that all the three quantities $C_{\delta}, C_{\epsilon}, C_{\mu}$ are positive for all $B>0$. Fig. \ref{f:C} illustrates the behavior of these quantities as a function of $B/B_{Q}$.

In the weak field limit, $B\ll B_{Q}$, these quantities take the known values
\begin{align}
C_{\delta} & =(2/45)\alpha (B/B_{Q})^{2},\label{cdweak}\\
C_{\epsilon} & =(4/45)\alpha (B/B_{Q})^{2},\label{ceweak}\\
C_{\mu} & =(7/45)\alpha (B/B_{Q})^{2}.\label{cmweak}
\end{align}

In contrast, is the very strong field limit, $B\gg B_{Q}$, the quantities scale as
\begin{align}
C_{\delta} & \propto \log(B/B_{Q}),\label{cdstrong}\\
C_{\epsilon} & \propto (B/B_{Q}),\label{cestrong}\\
C_{\mu} & \sim const.\label{cmstrong}
\end{align}

\section{QED plasma framework}
\label{s:framework}

The system of equations (\ref{m1},\ref{m2}), (\ref{gauss-my},\ref{ampere-my}) represents the full system of nonlinear Maxwell's equations. These equations are valid for a system with an arbitrarily strong magnetic field but is limited to the case of the vanishingly small electric field, compared to the critical field. In the presence of plasma, the system should be supplemented with the equations representing the plasma response via the source terms $\rho$ and ${\bf j}$. The actual form of the plasma response depends on the problem at hand. One can, for instance, calculate these quantities from a particle distribution function, if it is known:
\begin{align}
\rho & =\sum_{s} q_{s}\int f_{s}\, d^{3}p,\\
{\bf j} & =\sum_{s} q_{s}\int {\bf v}f_{s}\, d^{3}p,
\end{align}
where summation is over all plasma species $s$ having charges $q_{s}$ and integration is over the 3-momenta. Note, this approach is good for theoretical studies and is also particularly useful in particle-in-cell (PIC) simulations. 

To proceed further with the development of the QED-plasma framework, which is to be used to explore linear plasma waves and instabilities in the super-strong magnetic fields, we should assume that ${\bf E}=\tilde{\bf E}$, ${\bf B}={\bf B}_0+\tilde{\bf B}$ with $\tilde{\bf E},\tilde{\bf B}\ll{\bf B}_0$. The ``tilde'' denotes a perturbation. The background magnetic field ${\bf B}_0$ is assumed homogeneous, for simplicity.

\subsection{QED vacuum polarization contribution}

The relations between the fields in Eqs. (\ref{d},\ref{h}) have a general structure:
\begin{align}
D_i&=\epsilon_{ij}E_j=\left(\delta_{ij}+\chi_{ij}^{\rm vac}\right)E_j,\\
B_i&=\mu_{ij}H_j=\left(\delta_{ij}+\xi_{ij}^{\rm vac}\right)H_j.
\end{align}
The symmetry of the system at hand dictates that there are only two tensors from which $\epsilon_{ij}$ and $\mu_{ij}$ can be constructed. These are the isotropic tensor $\delta_{ij}$ (the Knocker-$\delta$) and the anisotropic tensor $b_{i}b_{j}$, where $b_i\equiv {B_0}_i/B_0$ is a unit vector in the direction of the background magnetic field (not to be mixed with $b=B/B_{Q}$).

\subsubsection{Weak field limit}

In the weak field, the electric and magnetic susceptibilities are well known \citep{landavshitz}
\begin{align}
\chi_{ij}^{\rm vac} &=C_\delta\left(-\delta_{ij}+\frac{7}{2}b_i b_j\right), \\
\xi_{ij}^{\rm vac} &=C_\delta\left(\delta_{ij}+2 b_i b_j\right), 
\end{align}
where $C_\delta$ is given by Eq. (\ref{cdweak}).

\subsubsection{$\mu_{ij}^{-1}$}

In what follows, we need the inverse $\mu_{ij}^{-1}$. It can only be constructed from tensors $\delta_{ij}$ and $b_i b_j$ too. Hence, one can write
\begin{equation}
\mu_{ij}=A\delta_{ij}+B b_ib_j, \qquad \mu_{ij}^{-1}=C\delta_{ij}+D b_ib_j,
\end{equation}
where $A,\ B,\ C,\ D$ are some scalar coefficients.
Upon contraction, $\mu_{ij}\mu_{jk}^{-1}=\delta_{ik}$, one obtains
\begin{equation}
C=\frac{1}{A},\qquad D=-\frac{B/A}{A+B}.
\end{equation}

It is convenient to write
\begin{equation}
\mu_{ij}^{-1}=\delta_{ij}+\eta_{ij}^{\rm vac}.
\end{equation}
Particularly, in the weak field limit
\begin{equation}
\eta_{ij}^{\rm vac}=-C_\delta\left(\delta_{ij}+2b_i b_j\right).
\end{equation}

\subsubsection{Arbitrarily strong field}

In the arbitrarily strong magnetic field, the equations for perturbed fields
\begin{align}
\tilde D_i&=\epsilon_{ij}\tilde E_j=\left(\delta_{ij}+\chi_{ij}^{\rm vac}\right)\tilde E_j,\\
\tilde H_i&=\mu_{ij}^{-1}\tilde B_j=\left(\delta_{ij}+\eta_{ij}^{\rm vac}\right)\tilde B_j
\end{align}
still hold and the vacuum susceptibilities are
\begin{align}
\chi_{ij}^{\rm vac}&=-C_{\delta}\delta_{ij}+C_{\epsilon}b_{i}b_{j},\label{chi}\\
\eta_{ij}^{\rm vac}&=-C_{\delta}\delta_{ij}-C_{\mu}b_{i}b_{j}.\label{eta}
\end{align}
The coefficients $C_\delta$, $C_\epsilon$, and $C_\mu$ are given by Eqs. (\ref{cd},\ref{ce},\ref{cm}). This can be directly obtained via linearization of Eqs. (\ref{gauss-my},\ref{ampere-my}). Note the simple physical meaning of these quantities. $C_\delta$ represents the {\em isotropic} contribution of QED vacuum to both susceptibilities, whereas $C_\epsilon$, and $C_\mu$ represent {\em anisotropic} contributions to the electric and magnetic susceptibilities, respectively.

\subsection{Plasma contribution}

In the presence of plasma, the current is related to the electric field via the complex-valued conductivity tensor
\begin{equation}
j_{i}=\sigma_{ij}\tilde E_{j},
\end{equation}
which leads to the straightforward redefinition of ${\bf D}$, namely $\partial_{t}{\bf D}\to \partial_{t}{\bf D}+4\pi {\bf j}$ in Maxwell's equations. We still can write $\tilde D_i=\epsilon_{ij}\tilde E_j$, but $\epsilon_{ij}$ takes the form
\begin{equation}
\epsilon_{ij}=\left.\delta_{ij}+\chi_{ij}^{\rm vac}+\chi_{ij}^{\rm plasma}\right. ,
\end{equation}
where $\chi_{ij}^{\rm plasma}$ is the plasma response. Assuming the perturbation is monochromatic of frequency $\omega$
\begin{equation}
\chi_{ij}^{\rm plasma}(\omega)=\frac{4\pi i \sigma_{ij}(\omega)}{\omega}.
\end{equation}

\subsection{QED-plasma framework}

For a monochromatic wave, $e^{-i\omega t+i {\bf k}\cdot {\bf x}}$, nonlinear Maxwell equations (\ref{m2},\ref{ampere-my}) become
\begin{align}
\varepsilon_{ijk} k_j \tilde E_k &= \frac{\omega}{c} \mu_{il} \tilde H_l,\\
\varepsilon_{ijk} k_j \tilde H_k &= -\frac{\omega}{c} \epsilon_{il} \tilde E_l.
\end{align}
The solution of the system of equations is the wave equation for a field:
\begin{equation}
\left(\frac{\omega^2}{c^2}\epsilon_{il}-\varepsilon_{ijk}\varepsilon_{lrq} k_j k_r\mu_{kq}^{-1}\right)\tilde E_l=0.
\end{equation}
The determinant of the matrix in the brackets is the dispersion relation, which solutions are eigenmodes of the system at hand. 

Using the identity 
\begin{align}
\varepsilon_{ijk}\varepsilon_{lrq}&=
\delta_{il}(\delta_{jm}\delta_{kn}-\delta_{jn}\delta_{km})
-\delta_{im}(\delta_{jl}\delta_{kn}-\delta_{jn}\delta_{kl})\nonumber\\
&\quad+\delta_{in}(\delta_{jl}\delta_{km}-\delta_{jm}\delta_{kl})
\end{align}
one obtains the dispersion equation for {\em arbitrary} 
$\epsilon_{ij}$ and $\mu_{ij}$
\begin{align}
&\textrm{det}\left[\frac{\omega^2}{c^2}\epsilon_{il}+\mu^{-1}_{il}k^{2}-\mu^{-1}\left(\delta_{il}k^2-k_i k_l\right)
\right. \nonumber\\
& \left.\quad\phantom{\frac{\omega^2}{c^2}}
+\delta_{il}\mu^{-1}_{jk}k_j k_k-\mu^{-1}_{ij}k_j k_l -\mu^{-1}_{lj}k_j k_i\right]=0
\end{align}
or, equivalently,
\begin{align}
&\textrm{det}\left[
\frac{\omega^2}{c^2}\left(\delta_{ij}+\chi_{ij}^{\rm vac}+ \chi_{ij}^{\rm plasma}\right)-\left(\delta_{ij}k^2-k_i k_j\right)\left(1+\eta^{\rm vac}\right)
\right.\nonumber\\
& \left.\phantom{\frac{\omega^2}{c^2}}
+\delta_{ij}\eta_{mn}^{\rm vac}k_m k_n+\eta^{\rm vac}_{ij}k^2-\left(\eta_{im}^{\rm vac} k_m k_j+\eta_{jm}^{\rm vac}k_m k_i\right)
\right]=0,
\label{disp-long}
\end{align}
where $\mu^{-1}\equiv \mu^{-1}_{ii}$ is the trace of $\mu^{-1}_{ij}$, $k^2=k_i k_i$ and $\eta^{\rm vac}=\eta_{ii}^{\rm vac}$.

Furthermore, assume the ambient field is homogeneous and is along $z$-axis, ${\bf B}=\left(0,~ 0,~ B\right)$. Hereafter we drop the subscript ``0'', for simplicity and ease of reading. Let us consider a wave propagating obliquely, at some angle $\theta$ with respect to the field so that ${\bf k}=\left(k \sin\theta,~ 0,~ k\cos\theta\right)$. Upon substitution of the above equations, the dispersion equation becomes
%\begin{strip}
\begin{widetext}
\begin{align}
&{\rm det}\left[\frac{\omega^2}{c^2}
\begin{pmatrix}
1-C_{\delta} & 0 & 0 \\
0 & 1-C_{\delta} & 0 \\
0 & 0 & 1-C_{\delta}+C_{\epsilon}
\end{pmatrix}\right.
+\frac{\omega^2}{c^2}\Bigg(\chi_{ij}^{\rm plasma}\Bigg)
\nonumber\\
&\qquad\left.-k^2\begin{pmatrix}
\cos^2\theta\left(1-C_{\delta}\right) & 0 & \sin\theta\cos\theta \left(1-C_{\delta}\right) \\
0 & \left(1-C_{\delta}-C_{\mu}\sin^2\theta\right) & 0 \\
\sin\theta\cos\theta\left(1-C_{\delta}\right) & 0 & \sin^2\theta\left(1-C_{\delta}\right)
\end{pmatrix}\right] =0.
\label{disp-main}
\end{align}
\end{widetext}
%\end{strip}

The derived equations (\ref{disp-long}) and (\ref{disp-main}) constitute the {\em QED-plasma framework} in its general form. They generalize the textbook plasma dispersion equation to the case of arbitrarily strong ambient magnetic fields, which enter the equations via vacuum susceptibilities. The plasma susceptibility is present in its most general form. In general, $\chi_{ij}^{\rm plasma}$ can be obtained directly from the plasma kinetic theory, e.g., from ``standard" plasma $\epsilon_{ij}=\delta_{ij}+\chi^{\rm plasma}_{ij}$. Here is the place to pick your model: composition, magnetization, plasma $\beta,\ T,\ n$, quantum statistics, and so on.

\section{Plasma example}

As an illustrative example of a magnetar plasma, let us consider a cold (so that pressure can be neglected and the sound speed vanishes), magnetized, $e^{\pm}$ plasma. Such a plasma is generally {\em non-neutral} because the Goldreich-Julian (GJ) density 
\begin{equation}
\rho_{GJ}=-\frac{{ \Omega_{NS}}\cdot {\bf B}}{2\pi c}
\end{equation}
depends on the relative orientation of the magnetic field and the neutron star spin, so it can be positive or negative, and it vanishes only then the spin and the magnetic field vectors are orthogonal. The GJ density is the minimum density of plasma in a magnetar magnetosphere. It is associated with the magnetosphere corotation. The associate GJ plasma density is of the order of $n_{GJ}\sim7\times10^{13}B_{15}P^{-1}{\rm cm}^{-3}$, where $B_{15}=B/(10^{15}\textrm{ gauss})$, $P$ is the period of the magnetar in seconds..

In magnetars, the magnetosphere can twisted by surface shear motions, so its field becomes nonpotential, $\nabla\times{\bf B}\not=0$. Therefore, it is threaded by electric currents \cite{magnetars-rev,BT07} with the current density, ${\bf j}=(c/4\pi)\nabla\times{\bf B}$. Thus, plasma density needed to maintain this current, $n=j/c e={\cal M}|\rho_{GJ}/e|$ exceeds the GJ density by a few orders of magnitude, where ${\cal M}\gg1$ is called the multiplicity. One can estimate that $n=j/c e\simeq 10^{17} B_{15}r_{6}^{-1}{\rm cm}^{-3}$, where $r_{6}=r/(10^{6}{\rm cm})$ is the radial distance from the center of the magnetar. 

Because of the very strong ambient field, the transverse kinetic energy of particles it radiated away very rapidly. Hence, plasma particles reside in the lowest Landau level. We assume that the plasma temperature small compared to the first excited Landau level, $kT_{\|}\ll \hbar\Omega$ (with $\Omega=e B/m_{e} c$ being the electron cyclotron frequency), so the transitions to excited levels are kinematically forbidden. We also consider the low-frequency modes only, $\omega<\Omega$, so the excitation of cyclotron harmonics would not be possible. This is reasonable, because at $B\sim B_{Q}$, the associated energy is comparable to the rest mass of the electrons/positrons, $\hbar\Omega\sim m_{e}c^{2}$. Therefore such processes might affect the plasma equilibrium values (e.g., its density), which we assume constant, for simplicity. As discussed above, the typical plasma densities are $n\sim10^{16}-10^{18}\textrm{ cm}^{-3}$. The corresponding typical plasma frequencies are $\omega_{p}\sim10^{12}-10^{13}\textrm{ s}^{-1}$. The characteristic cyclotron frequency is many orders of magnitude higher, $\Omega\sim10^{22}(B/B_{Q})\textrm{ s}^{-1}$. Thus, for the rest of the paper, we assume the ordering $\omega_{p}\ll\Omega$. Finally, we assume small occupation numbers, so the particle statistics (Bose/Fermi) is irrelevant. Quantum statistics in an $e^{\pm}$ plasma becomes important when the particle separation is smaller or of the order of the electron Compton wavelength, $d\sim1/n_{q}^{1/3}\lesssim\lambda_{e}\equiv\hbar/m_{e}c=3.9\times10^{-11}$~cm, that is, at plasma densities $n\gtrsim n_{q}=1.7\times10^{31}\textrm{ cm}^{-3}$.

\subsection{Plasma susceptibility}

The general expressions for $\epsilon_{ij}$ in a cold, multi-species plasma in a magnetic field are well known \citep{Aleksandrov-book,Akhiezer-book,DGbook}
\begin{equation}
\epsilon_{ij}^{\rm plasma}=\begin{pmatrix}
\epsilon_\bot & i\,g & 0 \\
-i\,g & \epsilon_\bot & 0 \\
0 & 0 & \epsilon_\| \\
\end{pmatrix},
\end{equation}
where
\begin{align}
\epsilon_\bot &= \epsilon_{xx}= \epsilon_{yy}= 1-\sum_s \frac{\omega_{ps}^2}{\omega^2-\Omega_s^2}, \\
\epsilon_\| &= \epsilon_{zz}= 1-\sum_s \frac{\omega_{ps}^2}{\omega^2}, \\
\epsilon_{xy} &=-\epsilon_{yx}=i\,g=-i\sum_s \frac{\omega_{ps}^2\Omega_s}{\omega\left(\omega^2-\Omega_s^2\right)}, \\
\epsilon_{xz}&= \epsilon_{zx}=\epsilon_{yz}=\epsilon_{zy}=0.
\end{align}
Here
\begin{align}
\omega_{ps}^2 &= \frac{4\pi q_s^2n_s}{m_s}, \\
\Omega_s &=\frac{q_s B}{m_s c}
\end{align}
are the plasma frequency and the cyclotron frequency of species $s$ (the sign of $\Omega_s$ takes the sign of the charge $q_{s}$).

For the electron-positron non-neutral plasma, one obtains
\begin{equation}
\chi_{ij}^{\rm plasma}=\begin{pmatrix}
\chi_\bot & i\,g & 0 \\
-i\,g & \chi_\bot & 0 \\
0 & 0 & \chi_\| \\
\end{pmatrix},
\end{equation} 
where
\begin{align}
\chi_\bot &= -\frac{\omega_{p}^2}{\omega^2-\Omega^2} ,\\
\chi_\| &= -\frac{\omega_{p}^2}{\omega^2}, \\
g&= -\frac{\omega_{p}^2\,\Omega}{\omega\left(\omega^2-\Omega^2\right)}\frac{\Delta n}{n}.
\label{g*}
\end{align}
Here, $e=|q_+|=|q_-|$,  $n=n_+ + n_-$ is the total density and $\Delta n= n_+ - n_-$. Thus, we also have
\begin{align}
\omega_{p}^2 &= \frac{4\pi e^2n}{m_e}, \\
\Omega &=\frac{e B}{m_e c}.
\end{align}

If we assume that $n_{GJ}$ represent the entirely non-neutral fraction of the plasma, then the ``non-neutrality fraction'', $\Delta n/n$, that will appear later can be approximated as $\Delta n/n\sim n_{GJ}/n\sim {\cal M}^{-1}$, i.e., as the inverse of the multiplicity factor.

\subsection{Dispersion equation}

\begin{figure*}
%\vskip0.5cm
\includegraphics[scale = 0.95]{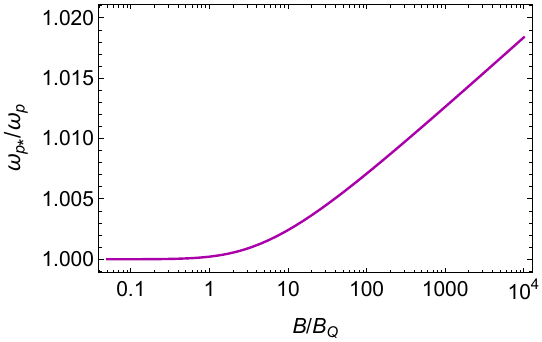}  %% file name without extension
\hskip0.3cm
\includegraphics[scale = 0.95]{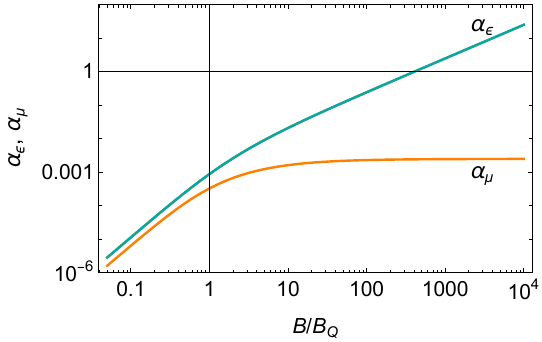}  %% file name without extension
\caption[]{QED modification of the plasma frequency (left panel) and quantities $\alpha_{\epsilon}, \alpha_{\mu}$ (right panel) as a function of $B/B_{Q}$.}
\label{f:alpha}
\end{figure*}

Let us now observe two things. First, Eq. (\ref{disp-long}) contains a common term $(1-C_{\delta})$. Second, all plasma susceptibilities $\chi_{ij}$ are proportional to $\omega_{p}^{2}$. One thus can renormalize the plasma frequency:
\begin{equation}
\omega_p^2\to\omega_{p*}^2\equiv \frac{\omega_p^2}{1-C_{\delta}}
\label{w*}
\end{equation}
and define 
\begin{align}
\alpha_\epsilon &=\frac{C_{\epsilon}}{1-C_{\delta}}, 
\label{alphae}\\
\alpha_\mu &=\frac{C_{\mu}}{1-C_{\delta}}.
\label{alpham}
\end{align}
Fig. \ref{f:alpha} illustrates the behavior of these parameters as a function of the field strength. 

With the above definitions, the dispersion relation is
%\begin{strip}
\begin{widetext}
\begin{equation}
{\rm det}\left[
\frac{\omega^2}{c^2}\begin{pmatrix}
\epsilon_{\bot*} & i\,g_* & 0 \\
-i\,g_* & \epsilon_{\bot*} & 0 \\
0 & 0 & \epsilon_{\|*} \\
\end{pmatrix}
-k^2\begin{pmatrix}
\cos^2\theta & 0 & \sin\theta\cos\theta \\
0 & 1-\alpha_\mu\sin^2\theta & 0 \\
\sin\theta\cos\theta & 0 & \sin^2\theta
\end{pmatrix}\right] =0,
\end{equation}
\end{widetext}
%\end{strip}
where
\begin{align}
\epsilon_{\bot*} &= 1-\frac{\omega_{p*}^2}{\omega^2-\Omega^2}, \\
\epsilon_{\|*} &= 1+\alpha_\epsilon-\frac{\omega_{p*}^2}{\omega^2}, \\
g_*&= -\frac{\omega_{p*}^2\,\Omega}{\omega\left(\omega^2-\Omega^2\right)}\frac{\Delta n}{n}.
\end{align}

Thus, the effect of the quantum vacuum reduces to: (i) renormalization of the plasma frequency and (ii) addition of two field-dependent coefficients to the dispersion relation: one in the $\omega^{2}$-term (proportional to $\alpha_\epsilon$) via $\epsilon_{\|*}$ and one in the $k^{2}$-term (proportional to $\alpha_\mu$) via an angle-dependent contribution. In extremely strong fields $B\gg B_{Q}$, (i) the only strong effect is due to $\alpha_\epsilon$, which grows linearly with the field strength $\alpha_\epsilon\propto B$, furthermore $\alpha_\epsilon$ exceeds unity $\alpha_\epsilon>1$ in the field $B/B_{Q}\gg 1/\alpha\sim 137$; (ii) the QED-modified plasma frequency grows logarithmically with the field strength $\omega_{p*}/\omega_{p}\simeq \omega_{p}(1+C_{\delta})\propto \log B$ and the relative change is of the order of percents; (iii), the contribution from $\alpha_\mu$ is small because it saturates at a value well below a percent, $\alpha_\mu\sim 10^{-3}$.

Introducing the index of refraction, $N^2=k^2 c^2/\omega^2$, we finally obtain 

%\begin{strip}
\begin{widetext}
\begin{equation}
{\rm det}
\begin{bmatrix}
N^2\cos^2\theta -\epsilon_{\bot*} & -i\,g_* & N^2\sin\theta\cos\theta \\
i\,g_* & N^2\left(1-\alpha_\mu\sin^2\theta\right)-\epsilon_{\bot*} & 0 \\
N^2\sin\theta\cos\theta & 0 & N^2\sin^2\theta-\epsilon_{\|*}
\end{bmatrix} = 0.
\end{equation}
\end{widetext}
%\end{strip}
Expansion of the determinant yields
\begin{equation}
N^4 A + N^2 B + C=0,
\label{N-disp}
\end{equation}
where the scalar coefficients, $A,\ B,\ C$, are 
\begin{align}
A&= \left(\epsilon_{\bot*}\sin^2\theta+\epsilon_{\|*}\cos^2\theta\right) 
\left(1-\alpha_\mu\sin^2\theta\right), 
\label{A}\\
B&= -\left[\epsilon_{\bot*}\,\epsilon_{\|*}
\left(1+\cos^2\theta-\alpha_\mu\sin^2\theta\right)
+\left(\epsilon_{\bot*}^2-g_*^2\right)\sin^2\theta
\right], 
\label{B}\\
C&= \epsilon_{\|*} \left(\epsilon_{\bot*}^2-g_*^2\right).
\label{C}
\end{align}
We use the same letter $B$ for one of the coefficients as for the field strength, hoping this would not cause any confusion.

\section{Analysis}
\label{s:analysis}

The full information on the dispersion curves is contained in Eq. (\ref{N-disp}). Its solution,
\begin{equation}
N^2=\frac{-B \pm\sqrt{B^{2}-4AC}}{2A},
\label{n2}
\end{equation}
is a fifth-order equation in $\omega^{2}$ for a given $k$. Hence it determines ten independent eigenfrequencies $\omega^{(\nu)}=\omega^{(\nu)}(k,\theta)$. Since each eigenfrequency $\omega^{(\nu)}$ has a conjugate $-\omega^{(\nu)}$, we will consider the eigenfrequencies to be positive-definite, for simplicity, and thus distinguish five branches of oscillations of a cold magnetized plasma. In a general case, Eq. (\ref{n2}) is analytically complicated. The explicit dispersion curves $\omega=\omega^{(\nu)}(k,\theta), \ \nu=1,2,3,4,5$ are not generally possible to find in a closed analytical form, except for some special cases. Therefore, we continue our analysis by exploring asymptotic regimes and finding characteristic frequencies, such as cutoffs and resonances.

\subsection{No plasma}

In the absence of plasma, two electromagnetic modes exist. 

First, when the fluctuating electric field of the wave is orthogonal to the ambient $B$-field, i.e., $\tilde E_y\not=0$ so that $\tilde{\bf E}\bot({\bf k},{\bf B})$-plane, the index of refraction is
\begin{equation}
N_\bot^2=\frac{1}{1-\alpha_\mu\sin^2\theta}
=\frac{1-C_\delta}{1-C_\delta+C_\mu\sin^2\theta}.
\label{N-perp}
\end{equation}
In the case of orthogonal propagation, ${\bf k}\bot {\bf B}$, this wave becomes the so-called extraordinary wave or ``X-mode''. Hence, one can call this wave for arbitrary $\theta$ the ``{\em oblique X-mode}''.

Second, when the electric field lies in the $({\bf k},{\bf B})$-plane, i.e., $\tilde E_x, \tilde E_z\not=0$ so that $\tilde{\bf E}\in({\bf k},{\bf B})$-plane, the index of refraction is 
\begin{equation}
N_\|^2=\frac{1+\alpha_\epsilon}{1+\alpha_\epsilon\cos^2\theta}
=\frac{1-C_\delta+C_{\epsilon}}{1-C_\delta+C_{\epsilon}\cos^2\theta}.
\label{N-para}
\end{equation}
In the case of orthogonal propagation, this wave becomes the so-called ordinary wave or ``O-mode''. Hence, one can refer to it the ``{\em oblique O-mode}''.

The dependence of the wave index of refraction on its polarization manifests birefringence of QED vacuum.

\subsection{Resonances, $\omega_\infty$}

At certain frequencies, the index of refraction diverges. These are {\em resonances}. At a resonance $N\to\infty$, so $A=0$ in Eq. (\ref{A}), which yields 
\begin{align}
\omega^4\left(1+\alpha_\epsilon\cos^2\theta\right)
&-\omega^2\left[\omega_{p*}^2+\Omega^2\left(1+\alpha_\epsilon\cos^2\theta\right)\right] \nonumber\\
& +\omega_{p*}^2\Omega^2\cos^2\theta = 0.
\end{align}
We renormalize the plasma frequency (again) as follows
\begin{equation}
\omega_{p*}^2\to
\bar\omega_{p*}^2 \equiv \frac{\omega_{p*}^2}{1+\alpha_\epsilon\cos^2\theta}
=\frac{\omega_{p}^2}{\left(1-C_\delta\right)\left(1+\alpha_\epsilon\cos^2\theta\right)}, 
\label{omegapbar}
\end{equation}
so, the above equation becomes
\begin{equation}
\omega^4-\omega^2\left(\bar\omega_{p*}^2+\Omega^2\right) 
+\bar\omega_{p*}^2\Omega^2\cos^2\theta = 0.
\end{equation}
the solutions are
\begin{equation}
\omega_\infty^2=\frac{1}{2}\left[
\left(\bar\omega_{p*}^2+\Omega^2\right)\pm
\sqrt{\left(\bar\omega_{p*}^2+\Omega^2\right)^2
-4\bar\omega_{p*}^2\Omega^2\cos^2\theta}
\right].
\label{res}
\end{equation}
We remind that in a NS/magnetar magnetosphere, $\omega_p\ll\Omega$, which we assume throughout the paper. The above equation yields two positive resonance frequencies ${\omega_\infty^{(1)}},\ {\omega_\infty^{(2)}}$. Below are a few asymptotic cases.

\subsubsection{Parallel propagation}

In the case of the parallel propagation $\theta=0$, the resonances are
\begin{align}
{\omega_\infty^{(1)}} &=\Omega,\\
{\omega_\infty^{(2)}} &=\bar\omega_{p*},
\end{align}
i.e., the cyclotron and  renormalized plasma frequencies. Note that $\bar\omega_{p*}$ is not just the renormalized plasma frequency $\omega_{p*}$, but an angle-dependent one, see Eq. (\ref{omegapbar}).

\subsubsection{Perpendicular propagation}

In the case of the perpendicular propagation $\theta=\pi/2$, the resonances are
\begin{align}
{\omega_\infty^{(1)}} &=\sqrt{\bar\omega_{p*}^2+\Omega^2},\\
{\omega_\infty^{(2)}} &=\frac{\bar\omega_{p*}\Omega}{\sqrt{\bar\omega_{p*}^2+\Omega^2}}\cos\theta\to0.
\label{winfty-Alfven}
\end{align}
That is, the upper resonance becomes hybrid and the lower one disappears as $\cos\theta$.

\subsubsection{Oblique propagation}

\begin{figure}
\centering
\includegraphics[scale = 0.95]{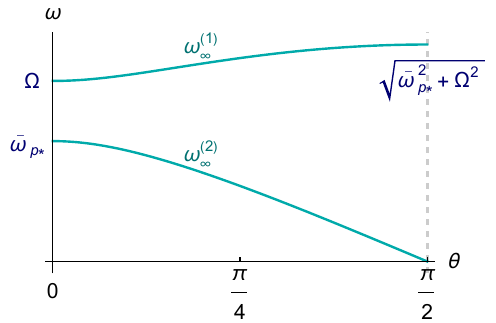}  %% file name without extension
\caption[]{The behavior of the two resonance frequencies, obtained from $A=0$ of Eq, (\ref{A}), i.e., Eq. (\ref{res}),  as a function of the wave propagation angle with respect to the ambient field. Note that $\bar\omega_{p*}$ differs from the renormalized plasma frequency, $\omega_{p*}$, and is given by Eq. (\ref{omegapbar}).
}
\label{f:res}
\end{figure}

The oblique propagation case is shown in Fig. \ref{f:res}.
%%%
%%%
It shows the behavior of the two resonance frequencies as a function of the propagation angle.

\subsubsection{Special case}

When $A\to0$ only one of the refraction indexes diverges, another one remains constant:
\begin{equation}
N^2=-C/B.
\end{equation}
A resonance on this branch is also possible if the coefficient $B\to0$ as well. In this case, the third resonance exists and is found from the condition $B=0$, which reads
\begin{equation}
\epsilon_{\bot*}\,\epsilon_{\|*}
\left(1+\cos^2\theta-\alpha_\mu\sin^2\theta\right)
+\left(\epsilon_{\bot*}^2-g_*^2\right)\sin^2\theta = 0.
\end{equation}
The general general solution of this equation, $\omega_{\infty}^{(3)}$, is analytically complicated.

All the resonances are illustrated in Fig. \ref{f:n2-disp-label}. The dependence on parameters is discussed in Section \ref{s:results}.

\subsection{Cutoffs, $\omega_0$}

At some frequencies, waves become non-propagating if their index of refraction becomes imaginary, $N^{2}<0$. Thus, the cutoff frequencies (or just {\em cutoffs}) are defined by $N=0$ equation. Thus, cutoffs are solutions of $C=0$ given by Eq. (\ref{C}):
\begin{equation}
\epsilon_{\|*} \left(\epsilon_{\bot*}^2-g_*^2\right)=0.
\end{equation}
This equation breaks up into two independent equations

The first one is 
\begin{equation}
\epsilon_{\|*} =0
\end{equation}
with the solution
\begin{equation}
{\omega_0^{(1)}}=\frac{\omega_{p*}}{\sqrt{1+\alpha_\epsilon}}.
\label{w0-Omode}
\end{equation}

The second equation is
\begin{equation}
\epsilon_{\bot*}^2-g_*^2=0,
\end{equation}
which reads
\begin{equation}
\omega^3-\omega\left(\omega_{p*}^2+\Omega^2\right)
\pm\omega_{p*}^2\Omega\frac{\Delta n}{n}=0.
\end{equation}

Its simple analytical solutions can be found in two limiting cases. 

First, $\left|{\Delta n}/{n}\right|\ll1$. Then the plasma is either (i) almost electrically neutral or (ii) the multiplicity of the electron-positron cascade in the magnetar magnetosphere is very high. Then 
\begin{align}
\omega_0^{(2)}&\approx \frac{\omega_{p*}^2\Omega}{\omega_{p*}^2+\Omega^2}\left|\frac{\Delta n}{n}\right|,\\
\omega_0^{(3)}&\approx \sqrt{\omega_{p*}^2+\Omega^2}.
\end{align}
Note that in the electrically neutral plasma $\Delta n=0$, one of the cutoffs vanishes, $\omega_0^{(2)}=0$.

Second, the opposite case is $\left|{\Delta n}/{n}\right|\to1$. This case corresponds to (i) the plasma being composed of a single species, electron or positron, or (ii) the plasma is electron-ion and ion contribution is insignificant due to their large inertia. Such a situation may occur in magnetars with an ``untwisted'' magnetosphere, where the plasma density approaches the GJ value, that is the plasma multiplicity approaches unity ${\cal M}\to 1$. In laser plasmas, it is hard to expect creation of highly non-neutral plasma, however. So this regime may only be relevant for fast processes, $\omega\gg\omega_{p, ion}$. Then 
\begin{equation}
\omega_0^{(2,3)} = \sqrt{\omega_{p*}^2+\frac{1}{4}\Omega^2}\mp\frac{1}{2}\Omega.
\end{equation}
This result coincides with the classical plasma, with the renormalized plasma frequency, i.e., with $\omega_{p}$ being replaced with $\omega_{p*}$,

All the cutoffs are illustrated in Fig. \ref{f:n2-disp-label}. The dependence on parameters is discussed in Section \ref{s:results}.

\subsection{Low-frequency asymptotic}

At low frequencies, $\omega\to0$, the asymptotic behavior has three distinct cases.

\subsubsection{$\Delta n=0,~\cos^2\theta\not=0$}

In neutral plasma, there are two oblique waves with the indexes of refraction:
\begin{align}
N^2_+ &=\frac{1+\omega_{p*}^2/\Omega^2}{\cos^2\theta},\\
N^2_- &=\frac{1+\omega_{p*}^2/\Omega^2}{1-\alpha_\mu\sin^2\theta}.
\end{align}
Recalling that $N=kc/\omega$, it immediately yields the following asymptotic:
\begin{align}
\omega_+ &\propto k_\|, \\
\omega_- &\propto k.
\end{align}
The first wave is the Alfven wave and the second one is the fast magnetosonic wave. That is
\begin{align}
\omega &=k_\|v_{A}, 
\label{va}\\
\omega &= k v_{f},
\label{vf}
\end{align}
where the Alfven and fast magnetosonic speeds are:
\begin{align}
v_{A} &=\frac{c}{\left(1+\omega_{p*}^2/\Omega^2\right)^{1/2}}, \\
v_{f} &=c\left(\frac{1-\alpha_\mu\sin^2\theta}{1+\omega_{p*}^2/\Omega^2}\right)^{1/2}.
\end{align}
Note that the fast mode speed becomes dependent on $\theta$ in QED vacuum. Note also that both speeds are modified by the QED effect via renormalization of the plasma frequency. The modification of the speeds are not very big, however, even in an extremely strong magnetic field.

\subsubsection{$\Delta n\not=0,~\cos^2\theta\not=0$}

In the non-neutral electron-positron plasma, there is one low-frequency ($\omega\to0$) oblique wave. It has the index of refraction
\begin{equation}
N^2=\frac{1}{\cos\theta\left(1-\alpha_\mu\sin^2\theta\right)}\frac{\omega_{p*}^2}{\omega\Omega}\frac{\left|\Delta n\right|}{n}.
\end{equation}
The asymptotic behavior is 
\begin{equation}
\omega\propto k\,k_\|,
\end{equation}
which corresponds to the whistler wave dispersion relation
\begin{equation}
\omega=kk_\| c^{2}\left(1-\alpha_\mu\sin^2\theta\right)\frac{\Omega}{\omega_{p*}^2}\frac{n}{\left|\Delta n\right|}.
\end{equation}

\subsubsection{$\Delta n\not=0,~\cos^2\theta=0$}

In the non-neutral plasma, there are no propagating modes with ${\bf k}\bot{\bf B}$. The index of refraction is
\begin{equation}
N^2=
-\frac{\displaystyle{\frac{\omega_{p*}^{4}}{\omega^2\Omega^2}\left(\frac{\Delta n}{n}\right)^2}}{\left(1+\displaystyle{\frac{\omega_{p*}^2}{\Omega^2}}\right)\left(1-\alpha_\mu
+\displaystyle{\frac{\omega_{p*}^2}{\Omega^2}\left(\frac{\Delta n}{n}\right)^2}\right)},
\end{equation}
that is $N^{2}<0$ unless 
\begin{equation}
\alpha_{\mu}>1+\frac{\omega_{p*}^2}{\Omega^2}\left(\frac{\Delta n}{n}\right)^2,
\end{equation}
which is false, since $\alpha_{\mu}\ll1$ for all reasonable values of $B$ (for $B\ll10^{180}B_{Q}$, i.e., until the logarithmically growing $C_{\delta}$ becomes large, $C_{\delta}\approx1$, hence $\alpha_{\mu}\to\infty$),\footnote{Likely, this estimate would not hold when higher-order in $\alpha$ multi-loop corrections are included since they may have different $B$-field asymptotic. } see Eq. (\ref{alpham}).

\subsection{High-frequency asymptotic}

In the $\omega\to\infty$ asymptotic, the plasma response becomes negligible and the waves become vacuum waves with the dispersion relations
\begin{equation}
\omega=kc/N_{\bot,\|}
\end{equation}
with $N_{\bot},N_{\|}$ given by Eqs. (\ref{N-perp},\ref{N-para}).

\begin{figure*}
\centering
\includegraphics[scale = 0.9]{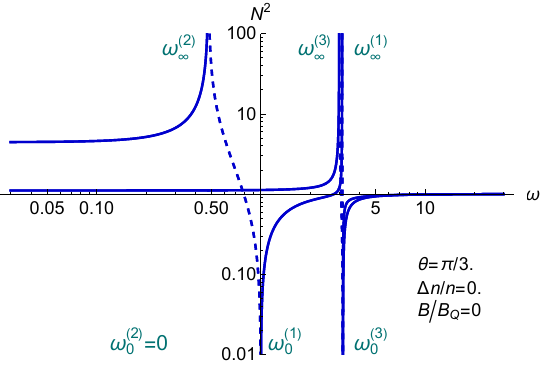}  %% file name without extension
\includegraphics[scale = 0.9]{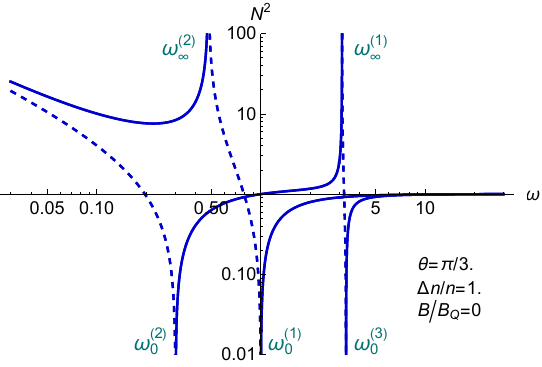}  %% file name without extension
\vskip0.3cm
\includegraphics[scale = 0.9]{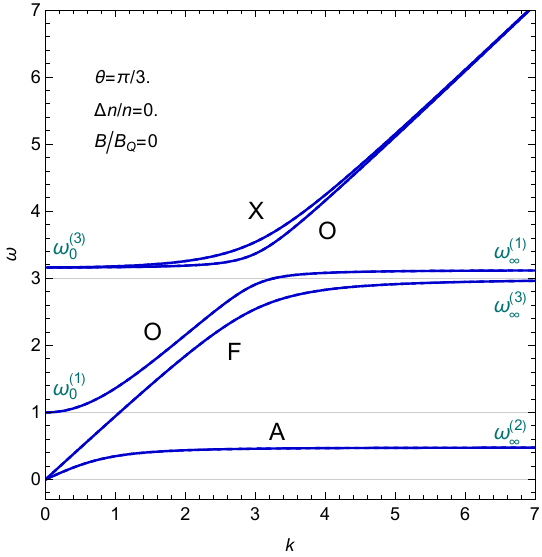}  %% file name without extension
\includegraphics[scale = 0.9]{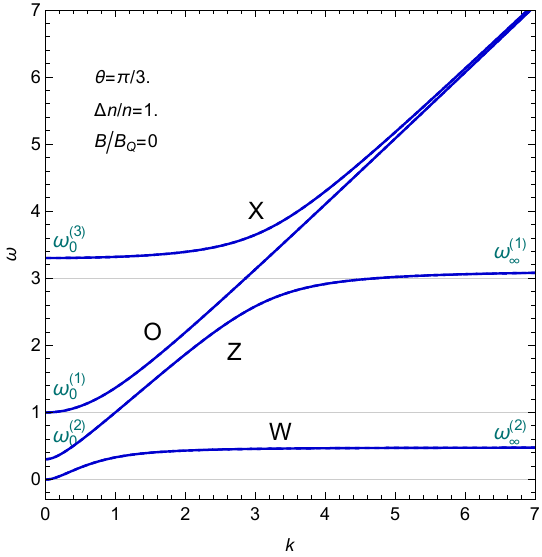} 
\caption[]{The schematic representation of the index of refraction squared $N^{2}(\omega)$ (top row) and the plasma dispersion curves $\omega(k)$ (bottom row) for the electrically neutral and non-neutral plasmas for $B\ll B_{Q}$. The units are arbitrary, but we set the speed of light to $c=1$. We set the numerical values of the plasma and cyclotron frequencies to be $\omega_{p}=1,\ \Omega=3$ and $\theta=\pi/3$. All resonances and cutoffs are labeled as in Sec. \ref{s:analysis}. Solid lines depict propagating waves, i.e., with $N^{2}>0$, and dashed lines depict evanescent branches with $N^{2}<0$. The wave branches are labeled as follows: ``A'' --- Alfven wave, ``F'' --- fast magnetosonic wave, ``X'' --- extraordinary electromagnetic wave, ``O'' --- ordinary electromagnetic wave (in a neutral plasma, it consists of two branches split around the cyclotron frequency), ``W'' --- whistler wave, ``Z'' --- Z-mode (the lower-frequency branch of the extraordinary wave, also called the slow extraordinary mode). 
}
\label{f:n2-disp-label}
\end{figure*}
\begin{figure*}
\centering
\includegraphics[scale = 0.75]{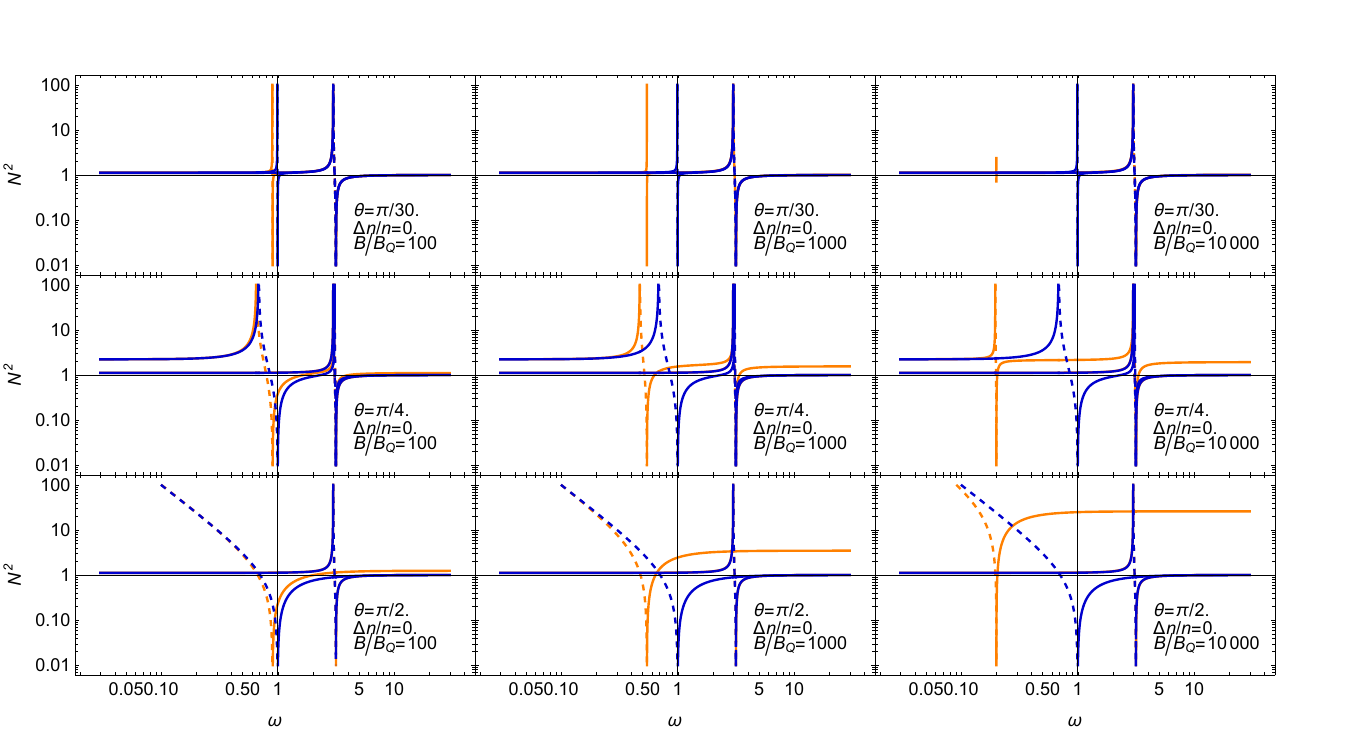}  %% file name without extension
\vskip-0.5cm
\includegraphics[scale = 0.75]{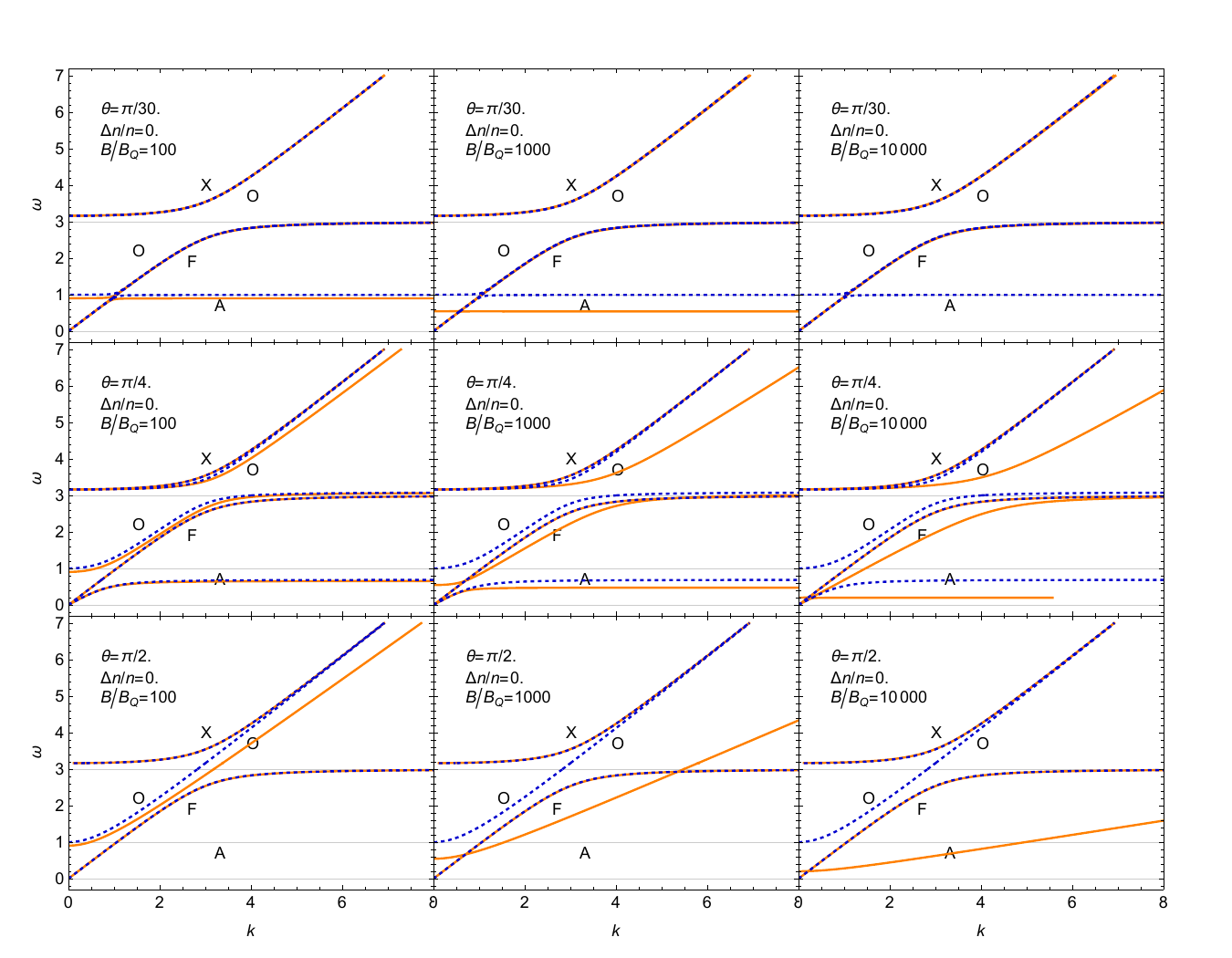}  %% file name without extension
\caption[]{The index of refraction $N^{2}(\omega)$ (top panel) and the plasma dispersion curves $\omega(k)$ (bottom panel) in an the electrically neutral, $\Delta n/n=0$, QED plasma as functions of the magnetic field, $B$, and the angle of propagation, $\theta$, with nearly parallel, oblique, and perpendicular propagation. The blue curves illustrate the non-QED regime and are shown for comparison. The plasma frequency is $\omega_{p}=1$ and the cyclotron frequency is $\Omega=3$. The latter is set to a constant, despite varying $B$, for the ease of comparison.  The wave branches are labeled as before. 
}
\label{f:n2-disp-set-deln0}
\end{figure*}
\begin{figure*}
\centering
\includegraphics[scale = 0.75]{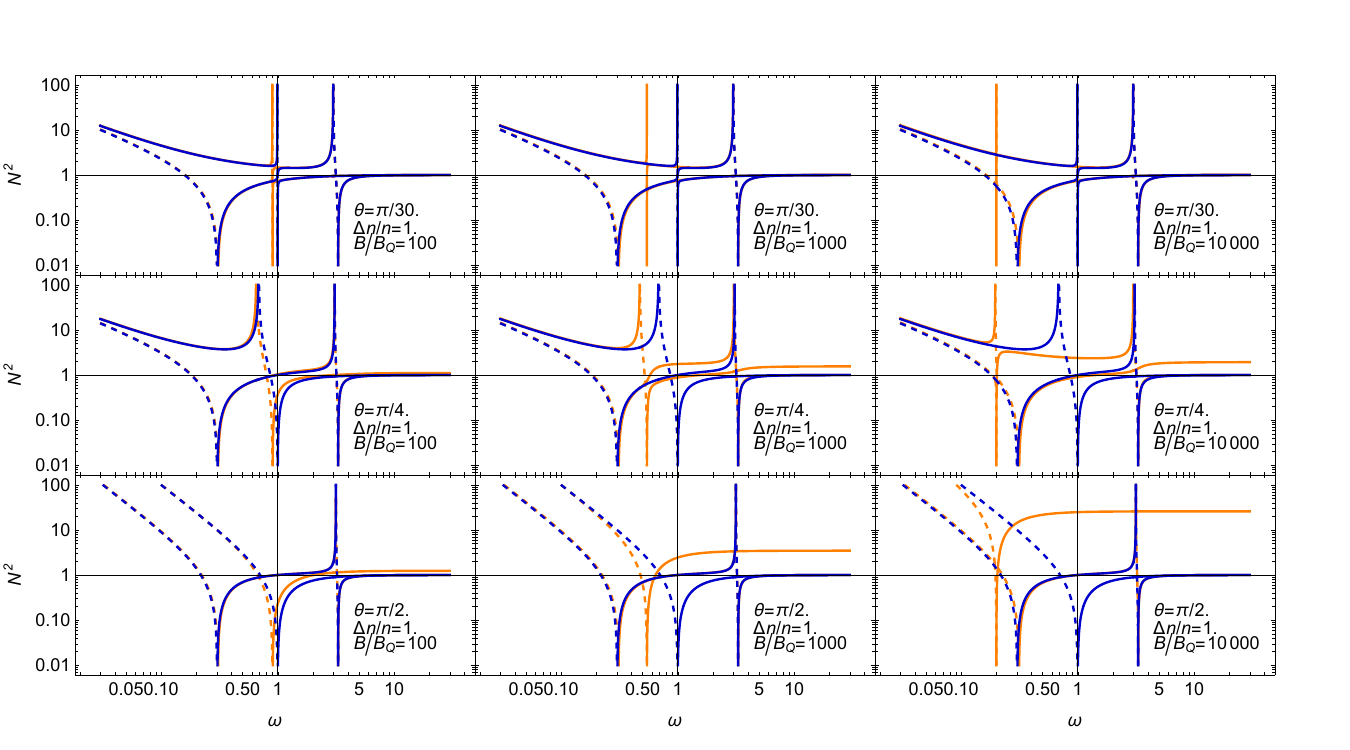}  %% file name without extension
\vskip-0.5cm
\includegraphics[scale = 0.75]{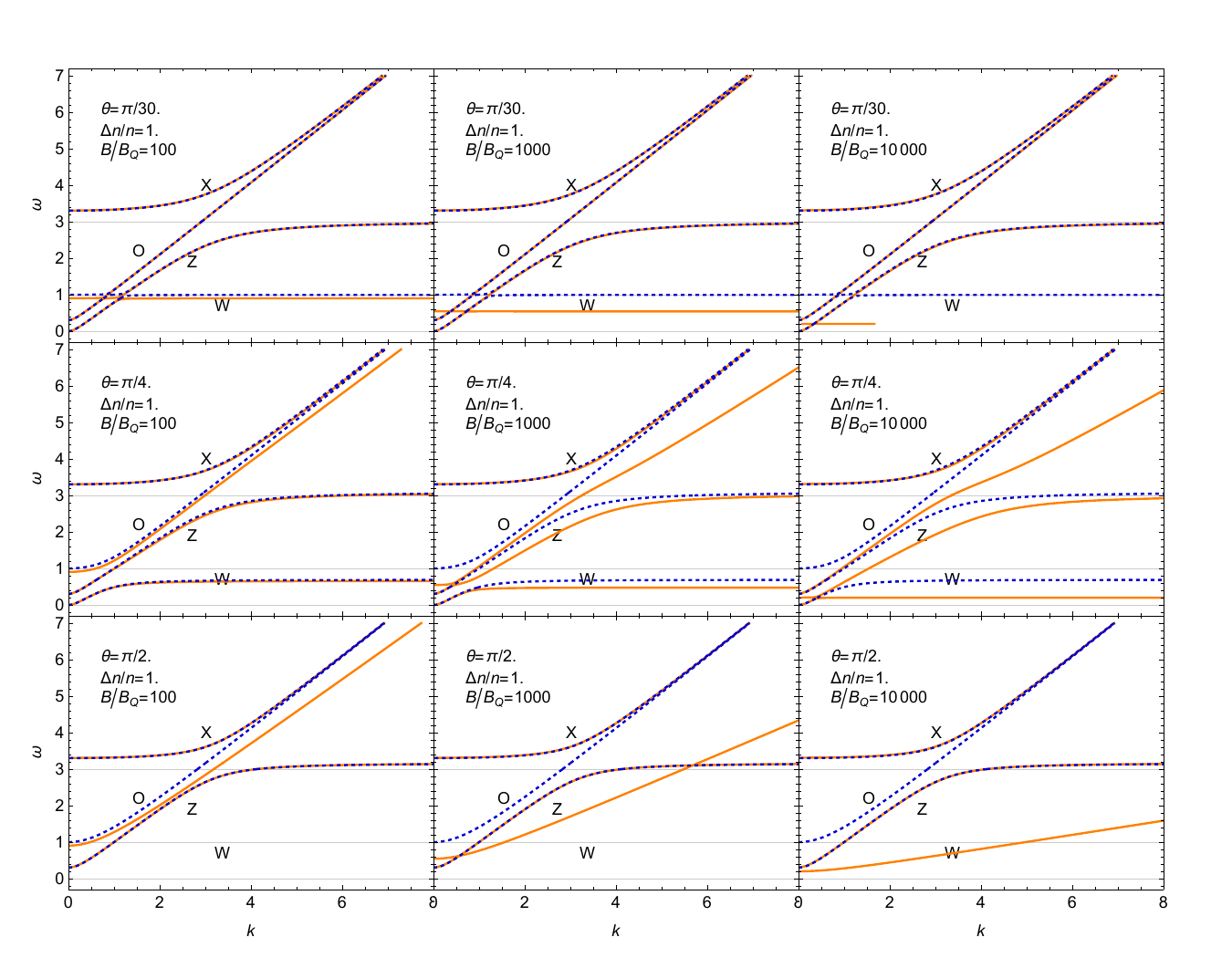}  %% file name without extension
\caption[]{The index of refraction $N^{2}(\omega)$ (top panel) and the plasma dispersion curves $\omega(k)$ (bottom panel) in an the electrically non-neutral, $\Delta n/n=1$, QED plasma as functions of the magnetic field, $B$, and the angle of propagation, $\theta$, with nearly parallel, oblique, and perpendicular propagation. The blue curves illustrate the non-QED regime and are shown for comparison. The plasma frequency is $\omega_{p}=1$ and the cyclotron frequency is $\Omega=3$. The latter is set to a constant, despite varying $B$, for the ease of comparison.  The wave branches are labeled as before. 
}
\label{f:n2-disp-set-deln1}
\end{figure*}
\begin{figure*}
\centering
\includegraphics[scale = 0.75]{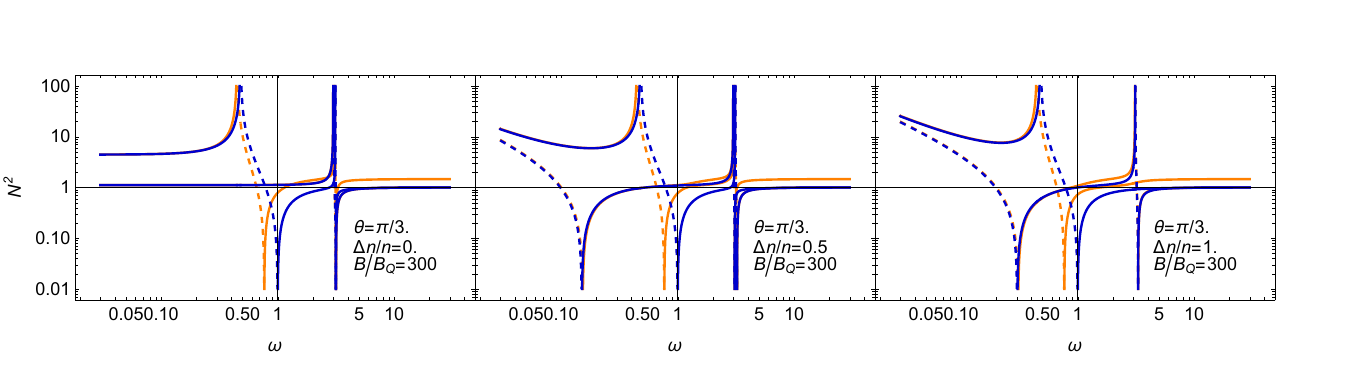}  %% file name without extension
\vskip-0.5cm
\includegraphics[scale = 0.75]{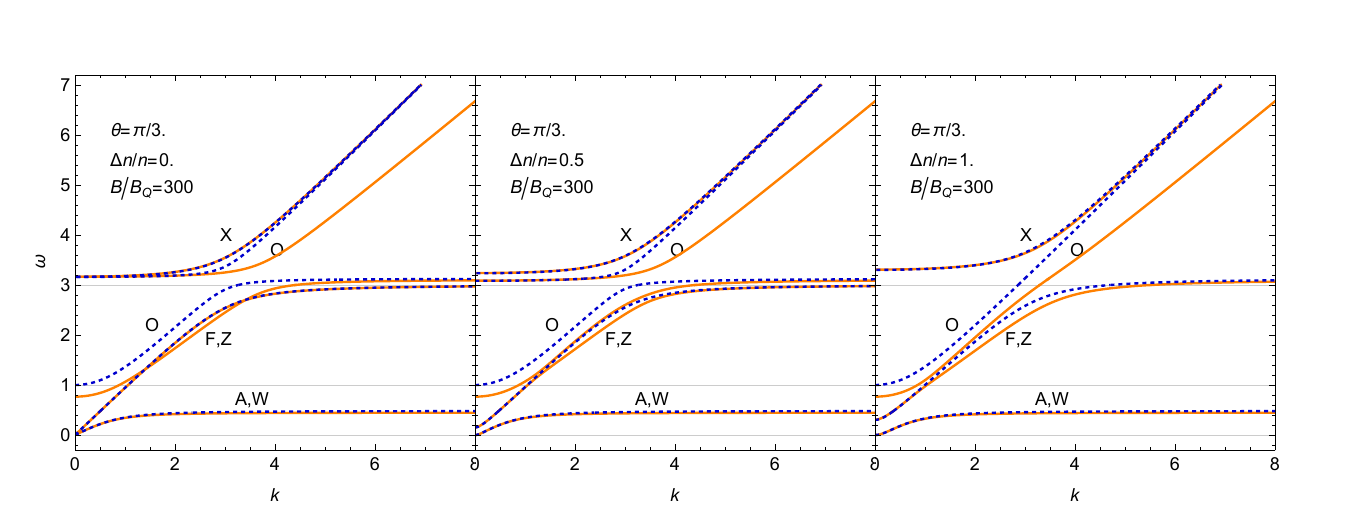}  %% file name without extension
\caption[]{The index of refraction $N^{2}(\omega)$ (top panel) and the plasma dispersion curves $\omega(k)$ (bottom panel) as function of the non-neutrality parameter, $\Delta n/n$. The plasma is in the QED regime with $B/B_{Q}=300$ and $\theta=\pi/3$. 
}
\label{f:n2-disp-set-b300}
\end{figure*}

\section{Numerical results}
\label{s:results}

Here we present numerical solutions of the full dispersion relation for various values of $B,\ \theta,\ \textrm{and}\ \Delta n$.

\subsection{Classical plasma}

The index of refraction $N^{2}(\omega)$ and the dispersion curves $\omega(k)$ in a classical (non-QED), cold, electron-positron plasma are shown in Fig. \ref{f:n2-disp-label}. The figure represents both electrically neutral and non-neutral plasmas in a very weak field, $B\ll B_{Q}$, when quantum effects are negligible. Dashed lines depict the plasma eigenmode branches with $N^{2}<0$. Solid lines depict propagating modes for which $N^{2}>0$. All resonances and cutoffs are labeled in accordance with the equations above in Sec. \ref{s:analysis}. For illustrative purposes, we chose the numerical values of the plasma and cyclotron frequencies to be $\omega_{p}=1,\ \Omega=3$ and the propagation angle is $\theta=\pi/3$. Note that in a realistic case of a magnetar magnetosphere, $\omega_{p}\ll \Omega$ by many orders of magnitude. The $\omega$ and $k$ units in the figure are arbitrary, but we set the speed of light to $c=1$. Since we chose $\omega_{p}=1$ it may seem that the frequencies are normalized by $\omega_{p}$ and $k$ are normalized by $\omega_{p}/c$, which is not so. Otherwise, $\Omega$ would be enormous and vary with $B$, which complicates understanding of the figure. 

The wave branches are labeled as follows: ``A'' is the Alfven wave ({\it a ka} shear Alfven wave), ``F'' is the fast magnetosonic wave ({\it a ka} compressible Alfven wave), ``X'' is the extraordinary oblique electromagnetic wave, ``O'' is the ordinary oblique electromagnetic wave (it consists of two branches in an neutral plasma, $\Delta n=0$, split near the cyclotron resonance, $\omega^{2}\sim\Omega^{2}$), ``W'' is the whistler wave, ``Z'' is the Z-mode (the lower-frequency branch of the extraordinary-like electromagnetic wave). These are all the ``classical'' modes in a non-QED, cold plasma. 

Fig. \ref{f:n2-disp-label} recovers the textbook result for the plasma modes in a cold, magnetized, electron-positron plasma. If the plasma is electrically neutral, $\Delta n=0$, there are two low-frequency modes: the Alfven and fast modes and the high-frequency ordinary and extraordinary electromagnetic waves; the O-mode is split into two branches near the cyclotron frequency, $\omega\sim \Omega$. We remind that at strictly perpendicular propagation, $\theta=\pi/2$, both modes are always linearly polarized. The X-mode has the electric field perpendicular to the background magnetic field and the O-mode has the wave electric field along $\bf B$. Both modes remain linearly polarized at oblique angles in an electrically neutral electron-positron plasma. It is clear from the fact that the dielectric tensor is diagonal in this case, i.e., $g_{*}=0$, see Eq. (\ref{g*}). 

In a non-neutral plasma, $\Delta n\not=0$, the modes are the whistler wave, O- and X-modes and the Z-mode, which can be considered the lower speed X-mode. This is clearly seen from the $N^{2}$ graph, where the two branches are connected by the $N^{2}<0$ segment, indicating that one mode is a continuation of another. At oblique angles and if the plasma non-neutrality parameter is $\Delta n/n<1$, the O-mode remains split at $\omega\sim \Omega$. However, when the plasma is composed of either electrons or positrons only, $|\Delta n/n|=1$, the resonance disappears, as is seen in Fig. \ref{f:n2-disp-label}. At oblique angles, $0<\theta<\pi/2$, both O- and X-modes are elliptically polarized, since $g_{*}\not=0$, Eq. (\ref{g*}).

%\vspace{1cm}
\subsection{QED plasma. Dependence on magnetic field and angle of propagation}

\subsubsection{Neutral plasma}

Fig. \ref{f:n2-disp-set-deln0} present the dispersion curves (solid orange cirves) for the electrically neutral plasma, $\Delta n=0$, in a super-strong $B$-field $10^{2}\le B/B_{Q}\le10^{4}$ at different angles of propagation. The dispersion curves in a non-QED plasma are shown with blue dashed curves, for comparison. 

First, in the case of parallel and quasi-parallel propagation, the fast and Alfven modes are degenerate at low frequencies, $\omega\to 0$. At intermediate frequencies, $\omega_{0}^{(1)}\lesssim\omega\lesssim\Omega$, the fast and O-mode are degenerate instead. At high frequencies, $\omega\gtrsim \Omega$, the O- and X-modes are degenerate since both have the index of refraction around unity. The QED-strong magnetic field is substantially reducing  the O-mode cutoff, $\omega_{0}^{(1)}$, and the Alfven resonance, $\omega_{\infty}^{(2)}$, frequencies. Thus, in extremely strong fields $B\ggg B_{Q}$, the fast mode and O-mode are practically always degenerate. 

Second, in the case of the perpendicular and quasi-perpendicular propagation, the Alfven mode disappears (the label in the figures are still present, for uniformity of the figure panel appearance). The X-mode and the fast mode dispersion curves are not appreciably affected (though quantitative changes do present, see Section \ref{s:analysis}). The O-mode has no cyclotron resonance. Its index of refraction is modified appreciably by the QED effects. This effect is nearly identical to the vacuum effect because at large frequencies, $\omega\gg\Omega$, the plasma dispersion is largely irrelevant because of plasma particle inertia. The indices of refraction for X- and O-modes are given by Eqs. (\ref{N-perp},\ref{N-para}).

Finally, the case of oblique propagation is intermediate between the above two cases. The most apparent QED effect is the change of the index of refraction of the O-mode, which is angle-dependent and greatly exceeds unity at $B\gtrsim10^{3}B_{Q}$. The substantial decrease of the Alfven resonance, $\omega_{\infty}^{(2)}$ at $k\to \infty$, is also noticeable. The QED corrections to the Alfven mode, fast-mode, and X-mode speeds are not considerable.

\subsubsection{Non-neutral plasma}

Fig. \ref{f:n2-disp-set-deln1} present the dispersion curves for the strongly non-neutral plasma, $|\Delta n/n|=1$, in a super-strong $B$-field $10^{2}\le B/B_{Q}\le10^{4}$ at different angles of propagation. The trends in this plasma are similar to those in the neutral plasma discussed above, with the substitution of fast mode for Z-mode and of Alfven mode for a whistler mode. In particular, the whistler resonance, $\omega_{\infty}^{(2)}$ at $k\to \infty$, decreases appreciably with increasing the ambient magnetic field strength. Also, the O-mode cutoff decreases with increasing $B$. The O-mode index of refraction increases with $B$ and approaches an asymptotic value, which depends on the angle of propagation only, $N_{\|}\to 1/\cos\theta$ at large $\omega$, as in vacuum, Eq. (\ref{N-para}). Other dispersion curves are modified only slightly in the QED regime. 

\textbf{}

\subsection{QED plasma. Dependence on non-neutrality}

Finally, we illustrate the role of plasma non-neutrality. Fig. \ref{f:n2-disp-set-b300} present the dispersion curves for
$B=300B_{Q}$, $\theta=\pi/3$, and $\Delta n/n=0,\ 0.5,\ 1$. We see that the global structure of the modes remains qualitatively the same even in a very strong $B$-field, compared to classical plasmas. Non-neutrality does change the mode structure, so that the Alfven mode with $\omega\propto k_{\|}$ dispersion takes the whistler dispersion, $\omega\propto kk_{\|}$, in non-neutral plasmas. Additionally, the O-mode becomes non-resonant near the cyclotron frequency when $|\Delta n/n|\to1$. The O-mode cutoff, $\omega_{0}^{(1)}$, is unaffected by plasma non-neutrality.

\section{Summary}
\label{s:concl}

\begin{figure*}
\centering
\includegraphics[scale = 0.89]{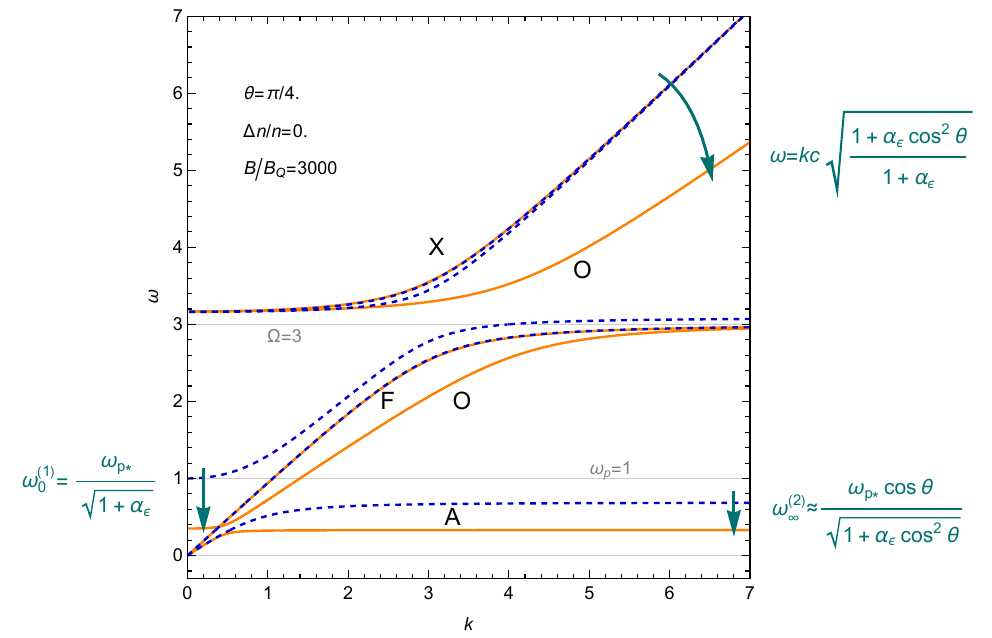}  %% file name without extension
\caption[]{Summary figure of plasma dispersion curves $\omega(k)$, showing salient features of QED effects on charge-neutral, $\Delta n=0$, cold electron-positron plasma. Other parameters are: $B=3000B_{Q}\sim10^{17}$~gauss, $\theta=\pi/4$, the plasma frequency is at $\omega_{p}=1$, and the cyclotron frequency is at $\Omega=3$. The wave branches are: ``A'' --- Alfven mode, ``F'' --- fast magnetosonic mode, ``X'' --- extraordinary electromagnetic mode, ``O'' --- ordinary electromagnetic mode. The most pronounced QED effects are shown by arrows. They are: (i) the $B$-field-induced transparency of the O-mode seen in the reduced wave cutoff frequency, $\omega_{0}^{(1)}$ as $k\to0$, (ii) the reduction of the phase speed of the O-mode at $\omega\gg\omega_{p*}$ (as in QED vacuum, too), and (iii) the reduction of the Alfvenic resonance frequency, $\omega_{\infty}^{(2)}$ as $k\to\infty$. The corresponding equations are also shown. Note, here the plasma frequency is also modified by QED, namely $\omega_{p*}=\omega_{p}/(1-C_{\delta})$. Dashed blue curves represent the dispersion curves in a classical (non-QED) plasma. 
}
\label{f:disp-sum}
\end{figure*}

Here we briefly summarize the results of this paper.

(I). We presented the Nonlinear Maxwell's equations, which result from QED-induced polarization and magnetization of vacuum. The equations are given in an implicit form, as well as explicitly in terms of the physical fields ${\bf E}$ and ${\bf B}$ and for an arbitrarily strong magnetic field. The first pair of Maxwell's equations is not affected by QED but merely represents a constraint on the fields, Eq. (\ref{m1},\ref{m2}). The second pair of Maxwell's equations, containing sources (charges and currents) does become modified, Eqs. (\ref{gauss},\ref{ampere}), as if the vacuum obtains polarization and magnetization, Eqs. (\ref{d},\ref{h}). The explicit form of Gauss' and Ampere's laws in an arbitrarily strong magnetic field are given by Eqs. (\ref{gauss-my},\ref{ampere-my}).

(II). We developed a {\em QED-plasma framework}, which allows systematic studies of linear plasma modes and instabilities in a QED-plasma with sub- and super-critical magnetic fields. The master equation is Eq. (\ref{disp-main}). It is written in a general form, independent of a specific plasma model, represented by the plasma electric susceptibility, $\chi_{ij}^{\rm plasma}$. Overall, the QED corrections to Maxwell's equations enter the general dispersion equation via three functions, dependent on the field strength, $C_{\delta},\ C_{\epsilon},\ C_{\mu}$, given by Eqs. (\ref{cd},\ref{ce},\ref{cm}), see also Fig \ref{f:C}. They have the following physical meaning. The vacuum electric and magnetic susceptibilities can be written as linear combinations of two available tensors: the isotropic tensor $\delta_{ij}$ and the anisotropic tensor associated with the magnetic field $b_{i}b_{j}$. The functions $C_{\delta},\ C_{\epsilon},\ C_{\mu}$ are the coefficients of these linear combinations, 
\begin{equation}
\chi_{ij}^{\rm vac},\eta_{ij}^{\rm vac}\propto -C_{\delta}\delta_{ij}\pm C_{\epsilon,\mu}b_{i}b_{j}. 
\end{equation}
Thus, $C_{\delta}$ represents the isotropic contribution to the vacuum susceptibilities, and $C_{\epsilon},\ C_{\mu}$ represent anisotropic contributions to the vacuum permittivity and permeability.

(III). We illustrated the use of the QED-plasma framework by considering a simple example of a cold, electron-positron plasma in a super-strong magnetic field. Both electrically neutral and non-neutral plasmas are considered. Such plasmas are expected to be present in the magnetospheres of magnetars. We also assumed that $\omega_{p}\ll\Omega$. The following results are obtained. They are illustrated in our `summary' figure, Fig. \ref{f:disp-sum}. 

(a). The qualitative picture of plasma eigenmodes in a super-critical field remains the same as in a classical plasma. The same five branches are present. No new modes appear. 

(b). The global QED effect, which appears throughout, is the renormalization of the plasma frequency, see Eq. (\ref{w*}) and Fig. \ref{f:alpha},
\begin{equation}
\omega_{p*}=\frac{\omega_{p}}{(1-C_{\delta})}.
\end{equation}
This effect is of the order of a percent, for typical magnetar fields. 

(c). Upon this re-normalization, the QED corrections enter the dispersion equations via only two functions $\alpha_{\epsilon},\ \alpha_{\mu}$ given by Eqs. (\ref{alphae},\ref{alpham}) and shown in Fig. \ref{f:alpha}. These two coefficients represent the QED contributions to the vacuum permittivity and permeability, respectively. One can see that the effect due to $\alpha_{\mu}$ is always below a percent, so it may be hard to measure. In contrast, the effect due to $\alpha_{\epsilon}$ can be large because it grows linearly with the field strength as long as $B\gg B_{Q}$. Note that the effect becomes appreciable when $\alpha_{\epsilon}\gtrsim1$, which corresponds to the fields $B\gtrsim 100B_{Q}\simeq 4\times 10^{15}$~gauss. 

(d). Although the global picture of the plasma modes in a super-critical QED-plasma remains similar to the classical plasma, there are numerous quantitative differences. The plasma cutoff and resonance frequencies, $\omega_{0}, \omega_{\infty}$, are greatly modified and become $B$-field-dependent and $\theta$-dependent. So do the low-frequency and high-frequency asymptotic scalings. All these are found analytically in Section \ref{s:analysis}. In particular, the Alfven and fast magnetosonic speeds are also quantitatively modified in a QED plasma, Eqs. (\ref{va},\ref{vf}).

(VI). The following QED-plasma effects are the most pronounced.
\begin{enumerate}
\item The {\em magnetically-induced transparency} of the ordinary mode. With increasing $B$-field strength, the O-mode can propagate at frequencies well below the plasma frequency. This is seen from the reduction of the O-mode cutoff frequency, as in Eq. (\ref{w0-Omode}),
\begin{equation}
\omega_{0}^{(1)}=\frac{\omega_{p*}}{\sqrt{1+\alpha_{\epsilon}}}.
\end{equation}
\item The {\em Alfven mode suppression} at high $\omega,\ k$. This is seen from the reduction of the Alfven resonance frequency, as in Eq. (\ref{winfty-Alfven}),
\begin{equation}
\omega_{\infty}^{(2)}\approx\frac{\omega_{p*}\cos\theta}{\sqrt{1+\alpha_{\epsilon}\cos^{2}\theta}}.
\end{equation}
\item The {\em slowdown} of the ordinary mode. This is seen from the increase of the index of refraction, Eq. (\ref{N-para}). The dispersion relation of the O-mode at high frequencies, $\omega\gg\omega_{p}$ is
\begin{equation}
\omega=kc\sqrt{\frac{1+\alpha_{\epsilon}\cos^{2}\theta}{1+\alpha_{\epsilon}}}.
\end{equation}
The effect depends on the angle of propagation of the wave with respect to the background $B$-field and vanished for the strictly parallel propagation. This effect does not require plasma to be present. It happens in vacuum too, and it is well known. 
\end{enumerate}

Fig. \ref{f:disp-sum} summarized these QED effects. Here, we show an example of the plasma dispersion curves (solid orange) in a neutral, cold, electron-positron plasma with $B\sim10^{17}$~gauss in the case of oblique propagation, $\theta=\pi/4$. These are compared with the non-QED plasma modes (dashed blue curves). Labeled are the Alfven (A), fast magnetosonic (F), ordinary (O) and extraordinary (X) modes. The trends with increasing $B$-field are shown by arrows. The plasma and cyclotron frequencies are equal to 1 and 3, respectively. All the enumerated QED effects are associated with the anisotropic vacuum contribution to the dielectric tensor, $\alpha_{\epsilon}$.

The QED-plasma effects above become significant or even dominant in super-critical magnetic fields, $B\gg B_{Q}$. They should be included in appropriate models and numerical simulations. These effects can be a `smoking gun' of a magnetic field of QED strength. Hopefully, they can be experimentally observed in laboratory laser plasma experiments and in observations of astrophysical sources, such as magnetars and gamma-ray burst progenitors.

%%%

\section*{Acknowledgements}

This research was supported by the National Science Foundation under Grant No. PHY-2010109. 
The author thanks Alexander Philippov for stimulating discussions and comments on the manuscript.

%% references
%%\bibliographystyle{aa}  %% aa.bst = astronomy standard name-year citations
%%\bibliographystyle{adsaa} %% adsaa.bst adds ADS bibcodes to references
%\bibliographystyle{mnras} 
%%\raggedright              %% only for adsaa with dvips, not for pdflatex
\bibliography{qed-plasmodes}    %% example.bib = bibtex entries copied from ADS

%%%%%%%%%%%%%%%%%%%%%%%%%%%%%%%%%%%%%%%%%%%%%%%%%%%%%%%%%%%%%%%%%%%%%%%%%%%%

\end{document}